\documentclass[prb]{revtex4-1} 

\usepackage{graphicx}
\usepackage{xcolor}
\usepackage[english]{babel}
\usepackage{amsmath}
\usepackage[colorlinks=true, allcolors=blue]{hyperref}
\usepackage{siunitx}
\usepackage[utf8]{inputenc}
\hypersetup{urlcolor=blue,
            linkcolor=blue,
            filecolor=blue }
\usepackage{url}

\begin{document}

\title{Regular and chaotic phase space fraction in the double pendulum}

\author{Santiago Cabrera$^1$, Edson D. Leonel$^2$, Arturo C. Marti$^1$}

\address{$^1$Instituto de Física, Universidad de la República, Igua 4225, 11400 Montevideo, Uruguay\\
$^2$Departamento de Física - Universidade Estadual Paulista - Unesp, Av.24A,
1515, Bela Vista, CEP, 13506-700 Rio Claro, SP, Brazil}

\begin{abstract}
The double coplanar pendulum is an example of the coexistence of regular and chaotic dynamics for equal energy values but different initial conditions. Regular trajectories predominate for low energies; as the energy is increased, the system passes through values where chaotic trajectories are abundant, and then, increasing the energy further, it is again dominated by regular trajectories. Given that the energetically accessible states are bounded, a relevant question is about  the  fraction of phase space regular or chaotic trajectories as the energy varies. In this paper, we calculate the relative abundance of chaotic trajectories in phase space, characterizing the trajectories using the maximum Lyapunov exponent, and find that, for low energies, it grows exponentially.
\end{abstract}

\maketitle

\section{Introduction}

\noindent 
In chaotic systems, instead of looking for long-term predictions about
the solutions, which are essentially impossible to obtain, it is
helpful to know the general properties of their evolution. An
essential feature of Hamiltonian dynamical systems is the coexistence
of regular and chaotic trajectories for different initial conditions
\cite{arnol2013mathematical,lichtenberg2013regular}. The non-trivial
combination of these trajectories gives rise to critical phenomena
such as dynamical traps \cite{zaslavsky2002dynamical} and anomalous
diffusion \cite{ishizaki1993anomalous}. One aspect that has been
little studied is the fraction of chaotic and regular trajectories as
we vary the system parameters. To answer this question meaningfully,
we must place ourselves in phase space where, thanks to Liouville's
theorem, the volume is conserved under canonical transformations and
especially during the time evolution given by the equations of motion
\cite{Goldstein}.

The double pendulum is a captivating and illustrative example within
chaos theory, showcasing the sensitivity to initial conditions
inherent in nonlinear dynamical systems. In the context of chaos, the
motion of a double pendulum is notoriously unpredictable, as slight
variations in the starting positions or initial velocities can lead to
vastly different trajectories over time. This sensitivity to initial
conditions is a hallmark of chaos where tiny perturbations can result
in significant deviations from the expected behavior. This chaotic
behavior results from the nonlinear equations governing the double
pendulum's motion.

Nonlinear descriptions, often formulated using the Lagrangian or
Hamiltonian formalism, are crucial in understanding the dynamics of
the double pendulum. The Lagrangian approach allows for a concise and
elegant representation of the system's kinetic and potential energy,
yielding the equations of motion through the principle of least
action. Alternatively, the Hamiltonian formalism provides insights
into the system's energy conservation and phase space dynamics,
offering a different perspective on the chaotic behavior of the double
pendulum. Researchers often leverage these formalisms to analyze the
intricate dynamics of chaotic systems, providing a foundation for
understanding chaotic phenomena in diverse fields.

The applicability of the double pendulum extends beyond mere
theoretical curiosity, finding relevance in various disciplines such
as physics \cite{luo2019period,dalessio2023}, engineering
\cite{laouina2022experimental}, and even in the study of biological
systems \cite{bazargan2014dynamics}. Engineers and physicists use
double pendulums as a testbed for chaotic behavior in mechanical
systems, helping to design and optimize systems that can tolerate and
control chaotic dynamics. By exploring chaotic dynamics in the double
pendulum, researchers gain valuable insights into the broader
implications of chaos theory and its potential applications in
real-world scenarios.

The distinction between regular and chaotic behavior is relevant to
the dynamical evolution of stellar systems.  In
Ref. \cite{muzzio2005spatial}, the abundance of regular and chaotic
orbits has been studied considering the non-zero Lyapunov exponent and
show that their spatial distributions substantially differ.  In
another approach, Manos and Athanssoula \cite{manos2011regular} found
that in a galactic model reduced to 2 degrees of freedom, the fraction
of chaotic trajectories is a non-monotonic function of the
energy. This information is vital in galaxy formation models, a
particularly active area in astrophysics, where the formation of
structures strongly depends on the orbits and morphology of the
celestial objects involved.

The planar double pendulum is a paradigmatic system with chaotic
dynamics \cite{shinbrot1992chaos,stachowiak2006numerical,Calvao_2015}.
In the low and high energy extremes, it possesses excellent regularity
in its motion - by theoretical predictions. There is an intermediate
energy range between them where essentially all initial conditions
result in chaotic motion.  This paper's main objective is the study of
the transition between regular and chaotic behavior from a global
perspective in the double-plane pendulum. With this aim, we calculate
the fraction of initial conditions corresponding to a section of the
phase space, in this case, the Poincaré section, that leads to a
chaotic evolution.  The results reported here concern a relevant and
paradigmatic system, such as the double pendulum, and could be
extended to various scenarios. The following section discusses the
general aspects of the double pendulum dynamics. We report the most
relevant results concerning the phase space fraction of chaotic
trajectories in Section~\ref{sec:experimentos}. Finally,
Section\ref{sec:conclusiones} provides a brief conclusion.

\section{The double pendulum}
\subsection{General description of the dynamics}
\label{ssec:general}
\begin{figure}
\centering
\includegraphics[scale=0.28]{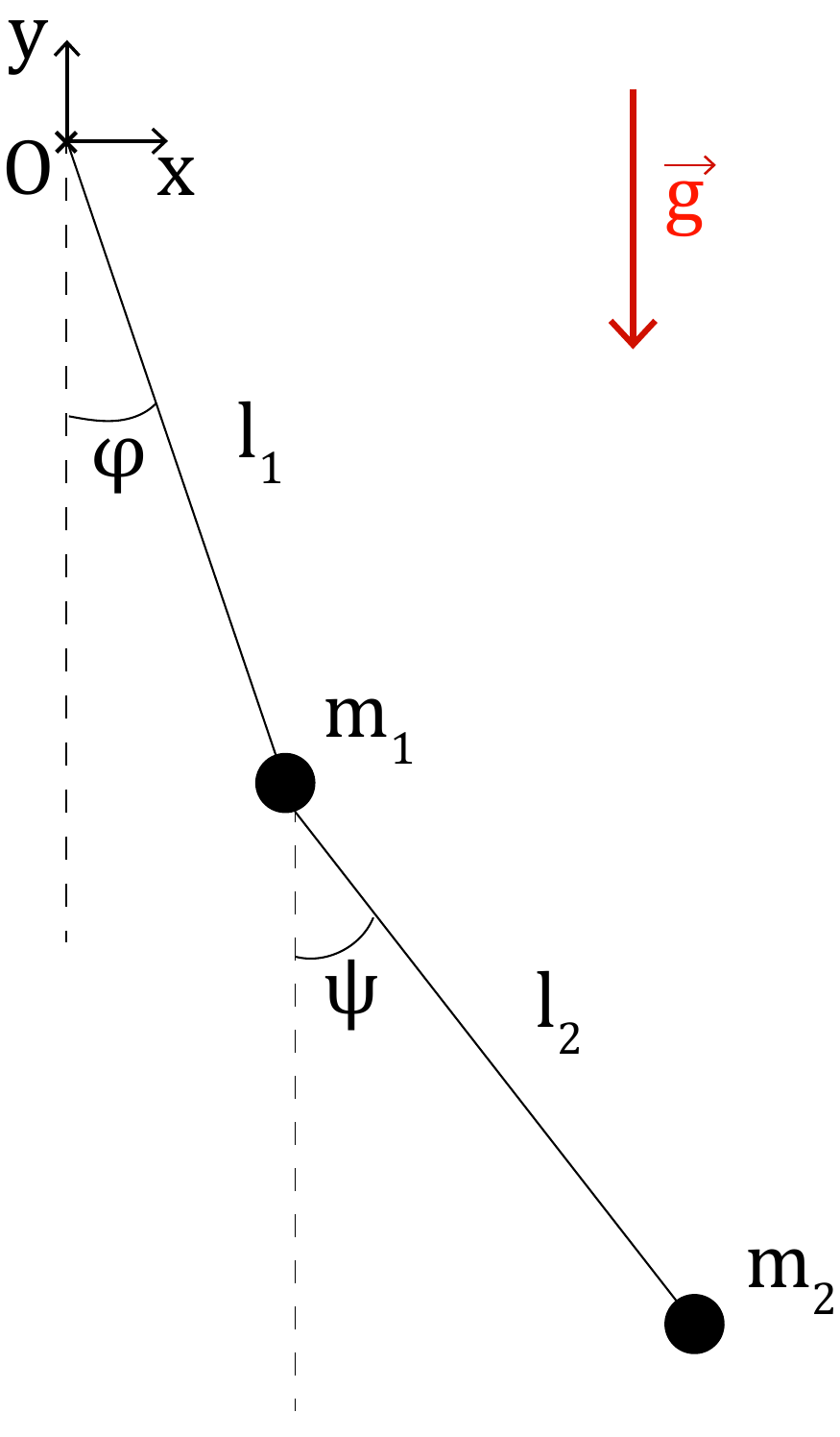}
\caption{Schematic diagram of the planar double pendulum.}
\label{fig:diagramaPendulo}
\end{figure}

A planar double pendulum is a physical system consisting of two
pendulums attached end-to-end, in which the second pendulum is
suspended from the lower end of the first. Each pendulum consists of
point masses, $m_1$ and $m_2$, attached to the end of massless
strings, lengths $l_1$ and $l_2$, allowing them to swing under the
influence of gravity, with gravitational acceleration $g$, in a single
vertical plane. As this system has two degrees of freedom and only one
conserved quantity, the mechanical energy, it exhibits complex and
chaotic behavior depending on the parameter values and the initial
conditions.  This system is often used in physics and mathematics to
illustrate chaotic and regular behavior from theoretical or
experimental perspectives \cite{shinbrot1992chaos}. In
Fig.~\ref{fig:diagramaPendulo} we show a system diagram where the
angles with respect to the vertical, $\varphi$, and $\psi$, are taken
as generalized coordinates.

The dynamics of this system can be analyzed using either the
Lagrangian or the Hamiltonian formalism. To begin with, we can express
the position of the masses in terms of $\varphi$ and $\psi$, as
$x_1=l_1\sin\varphi$ and $y_1=-l_1\cos\varphi$, and
$x_2=l_1\sin\varphi +l_2 \sin \psi$, $ y_2=-l_1\cos\varphi -l_2 \cos
\psi$.  The Lagrangian can be written as
\begin{equation}
L= \frac{1}{2}m_1l_1^2\dot\varphi^2+
\frac{1}{2}m_2(l_1^2\dot\varphi^2+l_2^2\dot\psi^2+2l_1l_2\dot\varphi\dot\psi\cos(\psi-\varphi))
-m_1 g l_1\cos\varphi-m_2g(l_1\cos\varphi+l_2\cos\psi).
\label{eq:lagrangiano}
\end{equation}
The equations of motion can be easily found from the Euler-Lagrange equations as
\begin{equation}
(m_1+m_2)l_1^2\ddot\varphi+m_2l_1l_2(\ddot\psi\cos(\psi-\varphi)-\dot\psi^2\sin(\psi-\varphi))    +(m_1+m_2)l_1g\sin\varphi =0
\label{eq:EuLa1}
\end{equation}
and
\begin{equation}
m_2l_2\ddot\psi+m_2l_1l_2(\ddot\varphi\cos(\psi-\varphi)\dot\varphi^2\sin(\psi-\varphi))+m_2l_2g\sin\psi=0
\label{eq:EuLa2}
\end{equation}

Throughout this work, we study the system's dynamics by analyzing the
Poincaré sections and calculating the Lyapunov characteristic
exponents described in the following Subsections. With these
objectives in mind, we integrate numerically the equations of motion
using a standard 4th-5th order Runge-Kutta method
\cite{MatLabode45}. The parameter values are $m_1=m_2=1$, $l_1=l_2=1$
and $g=9.81$.  Thanks to the numerical integration, we obtain the
temporal evolution of the generalized coordinates and velocities of
$\varphi(t)$, $\psi(t)$, $\dot\varphi(t)$ and $\dot\psi(t)$.  Although
it is simpler to work with the Lagrangian formalism since we are
interested in the fraction of chaotic trajectories in phase space, we
must also calculate the generalized momenta
\begin{equation}
    p_i=\frac{\partial L}{\partial \dot q_i}.
    \label{eq:definicionMomCan}
\end{equation}
Then, from the numerical integration of the generalized coordinates
and momenta, we obtain the momenta expressed as
$p_i(t)=p_i(q_i(t),\dot q_i(t))$.  This aspect is essential for
finding the Lyapunov exponents and the fraction of chaotic
trajectories defined in the phase space where Hamiltonian mechanics
occurs.

To conclude this summary, two limited dynamical behaviors are worth
mentioning. The first case is the limit of small oscillations. In that
case, the equations of motion reduce to a pair of easily integrable
second-order linear equations. Therefore, the orbits are regular, and
according to the ratio of the normal modes frequencies $\omega_1$ and
$\omega_2$, the generic motion will be periodic or quasiperiodic. This
limit corresponds to a low-energy case where the system does not move
too far from its equilibrium position. The null energy $E=0$
corresponds to both pendulums in the equilibrium position with
velocities equal to zero. Let us note that by slightly increasing the
energy so that the motions do not move too far away from the
equilibrium position, the equations of motion are no longer linear but
remain integrable. That is, the system describes non-harmonic
oscillations.  At the other end of the spectrum, the gravitational
potential energy becomes negligible for high energies compared to the
kinetic energy, and the perpendicular component of the angular
momentum is approximately conserved.  In this limit, the system's
dynamics are composed of ordered trajectories again. In short, the
double pendulum presents ordered dynamics in the low and high energy
limits. In contrast, in an intermediate range of energies, its
trajectories through the phase space are mainly chaotic. This work's
primary focus is verifying this prediction and quantitatively
analyzing the order-chaos-order transition.

\subsection{Poincaré sections}
\label{ssec:teoricoSecPoin}
The double pendulum has two degrees of freedom and one conserved
quantity, so the mechanical energy and dynamics occur in a
three-dimensional hypersurface. Following the criterion used by Korsch
et al. \cite{Chaos}, we consider two-dimensional Poincaré sections
corresponding to the points where the outer pendulum crosses the
vertical in a counter-clockwise direction
\begin{equation}
\psi=0 \;\;\; \text{and}\;\; \; l_2\dot\psi+l_1\dot\varphi\cos\varphi>0.  
\label{eq:definicionSecPoin}
\end{equation}

\begin{figure}[h]
\centering
\includegraphics[scale=0.6]{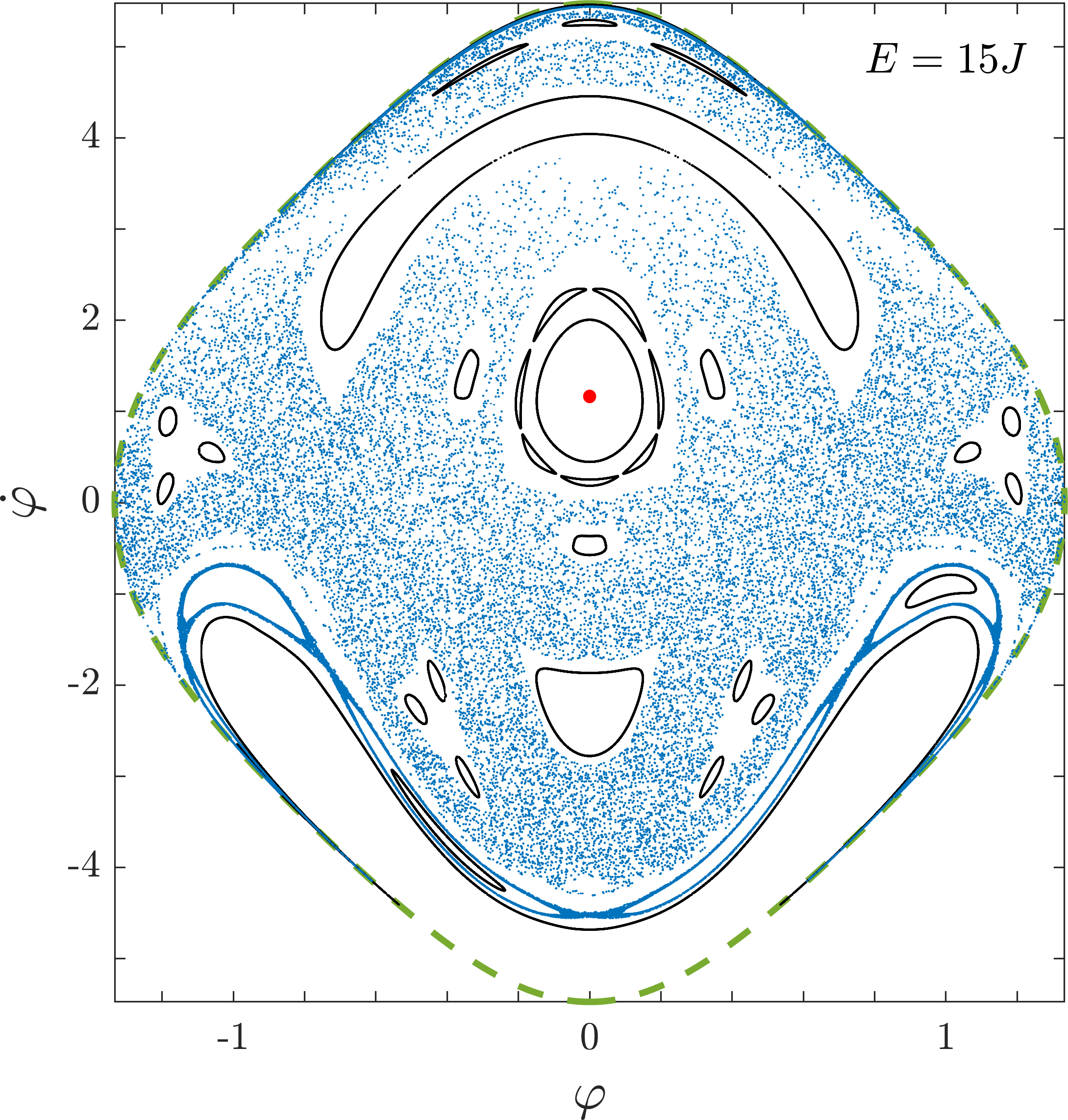}
\includegraphics[scale=0.3]{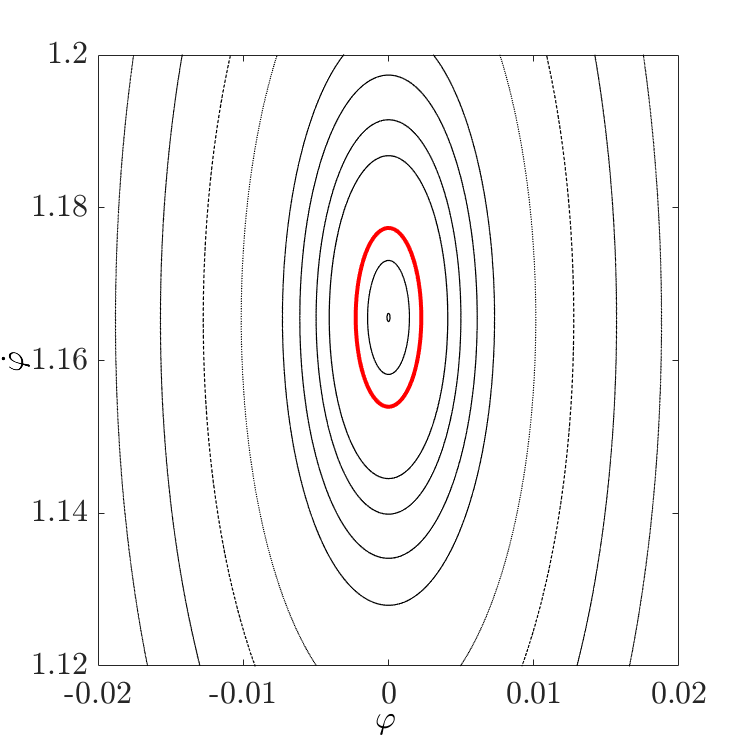}
\caption{Poincaré section for an energy value ($E=15J$) with
  coexistence of different dynamics. The left panel shows a global
  view where the dashed green curve represents the boundary of the
  energetically accessible region. Periodic motion is exemplified by
  the red point corresponding to orbits crossing the section at a
  single point.  The right panel shows an enlargement of the region
  surrounding the red point. Black orbits whose intersections form
  continuous curves correspond to quasiperiodic dynamics, and the
  orbits whose intersections fill non-zero areas (blue regions)
  correspond to chaotic motions. The chaotic trajectories were
  integrated for $10^4$s, integrating the equations of motion for
  longer times would lead to additional filling in the blue areas.  }
    \label{fig:Poincare}
\end{figure}

Depending on the value of the mechanical energy, the system exhibits
different behaviors that can be appreciated using Poincaré
sections. We can see in Fig.~\ref{fig:Poincare} three characteristic
dynamical behaviors for an energy $E=15J$
\begin{itemize}
    \item Finite sets of intersections. They correspond to periodic
      motions: the trajectory crosses the section through the same
      points after a time, for example, the red point in the center of
      the stable island that we find towards the center of the
      Poincaré section.
    \item Curve filling points (invariant curves) associated with
      quasiperiodic dynamics: the motion is regular but never passes
      through the same point twice (represented in black). 
    \item Area filling points, associated with chaotic trajectories:
      the trajectory crosses the Poincaré section erratically, without
      following any regularity or crossing the same threshold twice
      (blue regions).
\end{itemize}
We also note that the energetically accessible states in the Poincaré section are bounded, indicated in this figure by a green dashed closed line.

\subsection{Lyapunov characteristic exponents}
\label{ssec:ExpoDeLyapu}

As a complement to the analysis of the trajectories using Poincaré
sections, we also consider the calculation of the Lyapunov
characteristic exponents. In a conservative mechanical system with two
degrees of freedom, there are two opposite pairs of Lyapunov
exponents. One of them, parallel to the trajectories, corresponds to
null values. Therefore, the spectrum of Lyapunov exponents is
$(-\lambda,0,0,\lambda)$ and can be characterized entirely by
$\lambda$, the maximum Lyapunov exponent (MLE).  To numerically
calculate the MLE, we use a popular algorithm based on a convergent
iterative calculation \cite{wolf1985determining} extensively used in
the literature \cite{Chaos,stachowiak2006numerical}. The MLE can take
positive values in the case of chaotic dynamics or a null value for
regular solutions, which include periodic or quasiperiodic behavior.
Typically, computing the MLE involves integrating the equations of
motion over extended time intervals. Due to the slow convergence,
discerning between a positive or zero MLE becomes a subtle process. In
Subsection~\ref{ssec:clasificacionLyapu}, we elaborate on the criteria
employed to make this crucial distinction.

\section{Phase space fraction of chaotic trajectories}
\label{sec:experimentos}

\subsection{Algorithm}
\label{sec:alg}

We can determine the Poincaré section fraction corresponding to
chaotic trajectories with the tools presented in the previous
sections. There are several techniques to approximate what percentage
of initial conditions result in the chaotic motion of the system. One
of the techniques \cite{Chaso:3Rotors} compares the area occupied by
the chaotic trajectories by connecting the points corresponding to
intersections of such orbits with the section forming triangles with a
Delaunay triangulation. In this work, we propose another technique
exploiting the tools we have introduced, which gives results
comparable to those of the article above in less computer time. Our
algorithm is inspired by the well-known Monte Carlo method. The
procedure involves the following steps:

\begin{enumerate}
\item For a given energy, we find the maximum and minimum values of  $\varphi$ and $\dot{\varphi}$. For large energies where the inner pendulum can make a full rotation, we consider the range $[-\pi,\pi]$ for $\varphi$.
\item We define a grid of initial conditions
  $(\varphi_0,\dot{\varphi}_0)$ in the region
  $[\varphi_{\text{min}},\varphi_{\text{max}}]\times[\dot{\varphi}_{\text{min}},\dot{\varphi}_{\text{max}}]$
\item For each initial condition, we take $\psi_0=0$, and calculate the corresponding  $\dot{\psi}_0$  which verifies  the Eq.~\eqref{eq:definicionSecPoin} using the equations of motion. 
If the points lie in the Poincaré section, there is a solution for
$\dot{\psi_0}$ because they belong to the energy hypersurface of the system. We then have a vector of initial conditions in the configurations space, which are converted to the phase space using the canonical momenta definitions. We thus obtain a set of canonical initial conditions for each grid point in the Poincaré section.
\item For each initial condition, we integrate the equation of motion for sufficiently long periods and then calculate the MLE. 
\item We determine whether the value found corresponds to a regular or a chaotic orbit according to the criterion exposed in Subsection~\ref{ssec:clasificacionLyapu}.
\item Finally, after integrating all the initial conditions for a
  given energy, we sum the number of chaotic trajectories and divide
  by the total initial conditions.
\end{enumerate}

\begin{figure*}
  \centering
\includegraphics[width=0.4\textwidth]{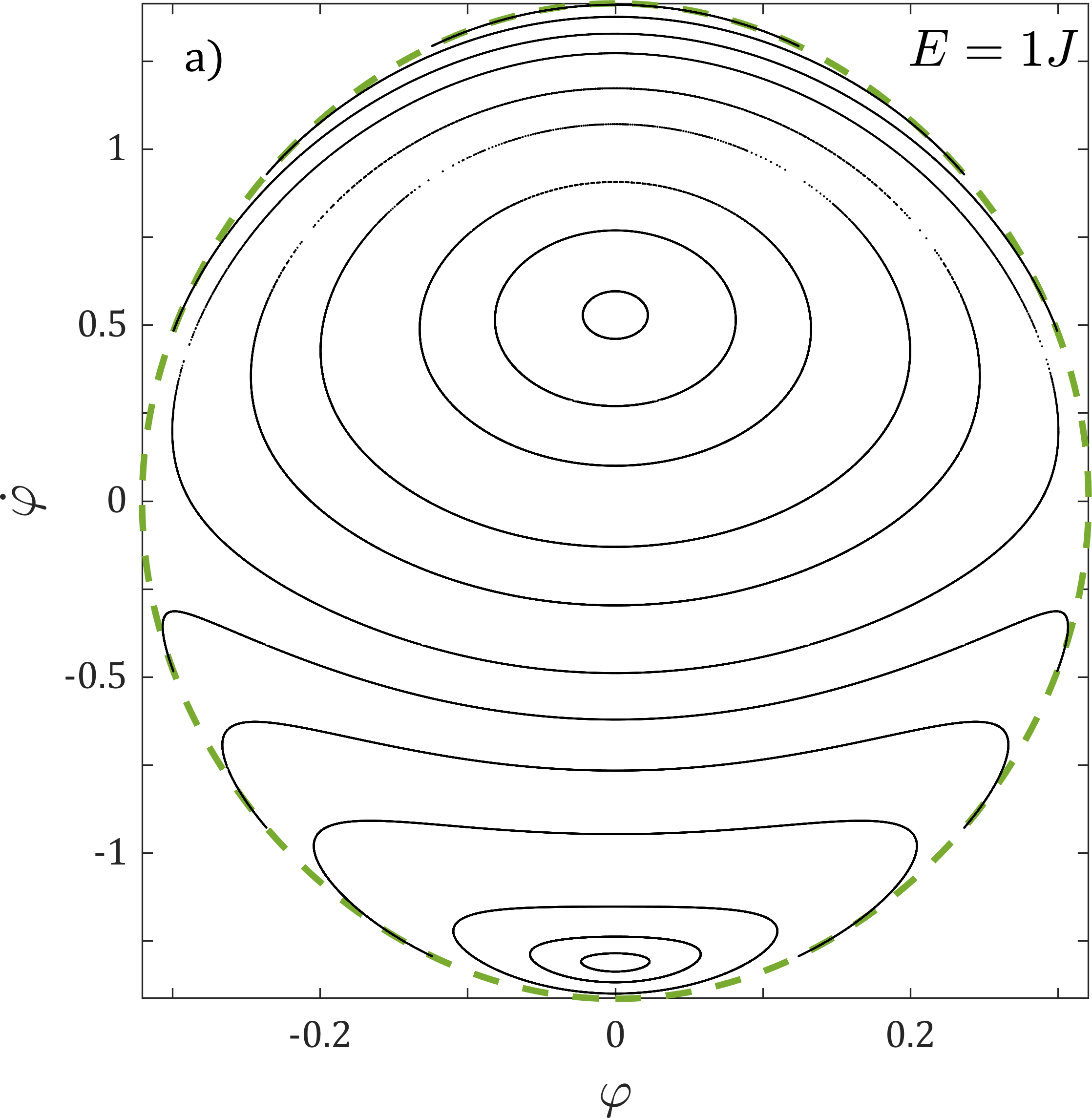}  
\includegraphics[width=0.39\textwidth]{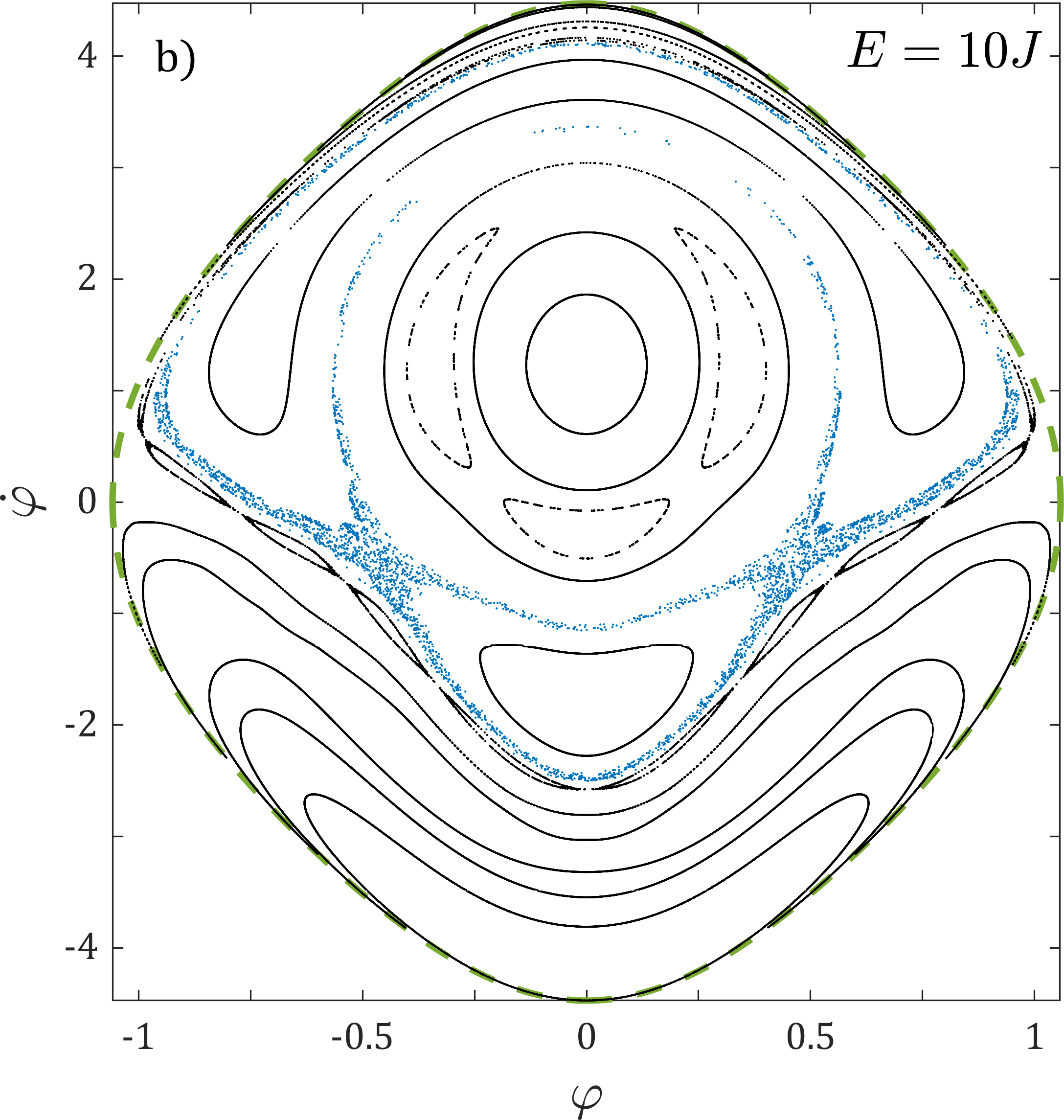}
\includegraphics[width=0.4\textwidth]{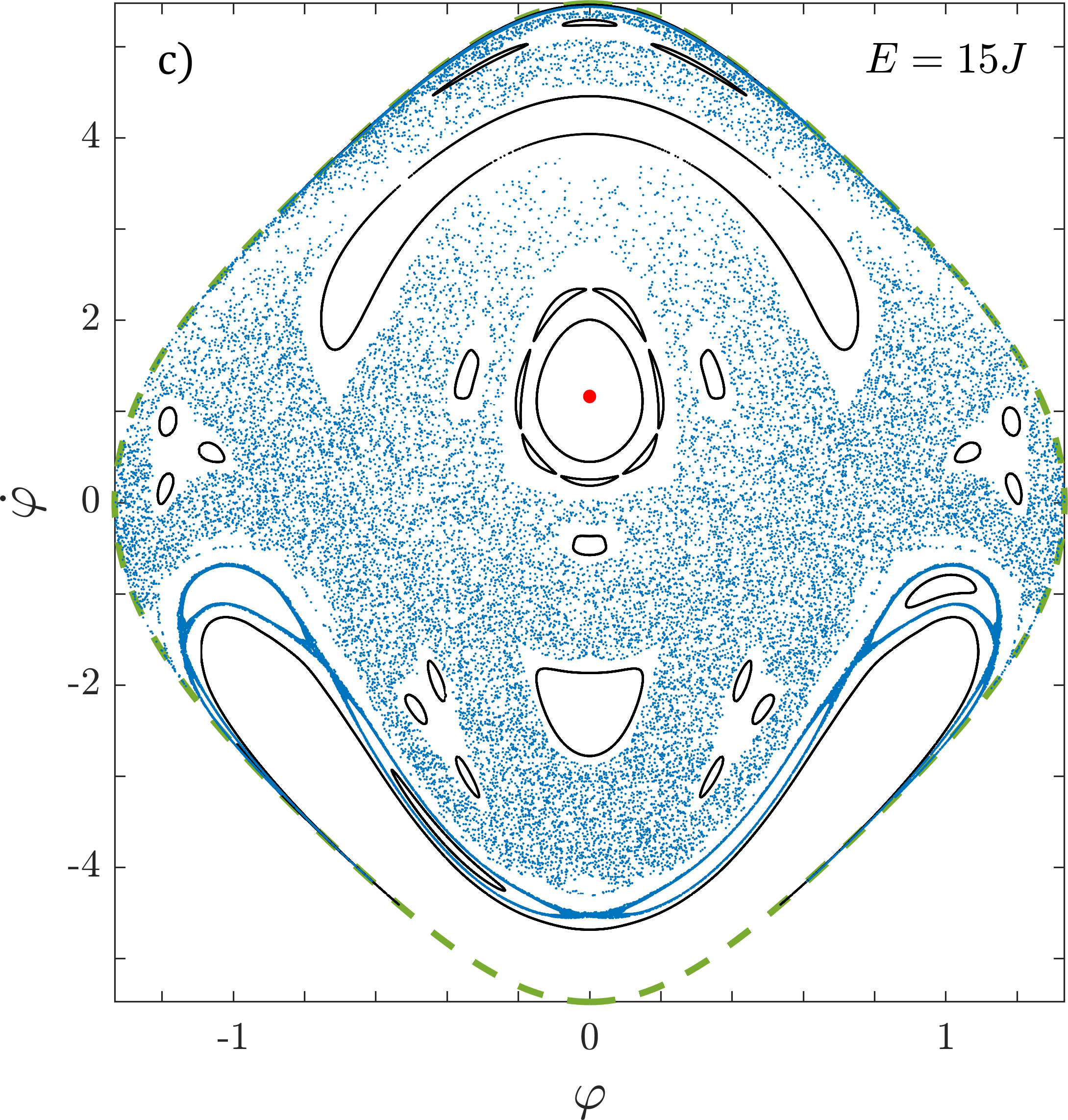}
\includegraphics[width=0.4\textwidth]{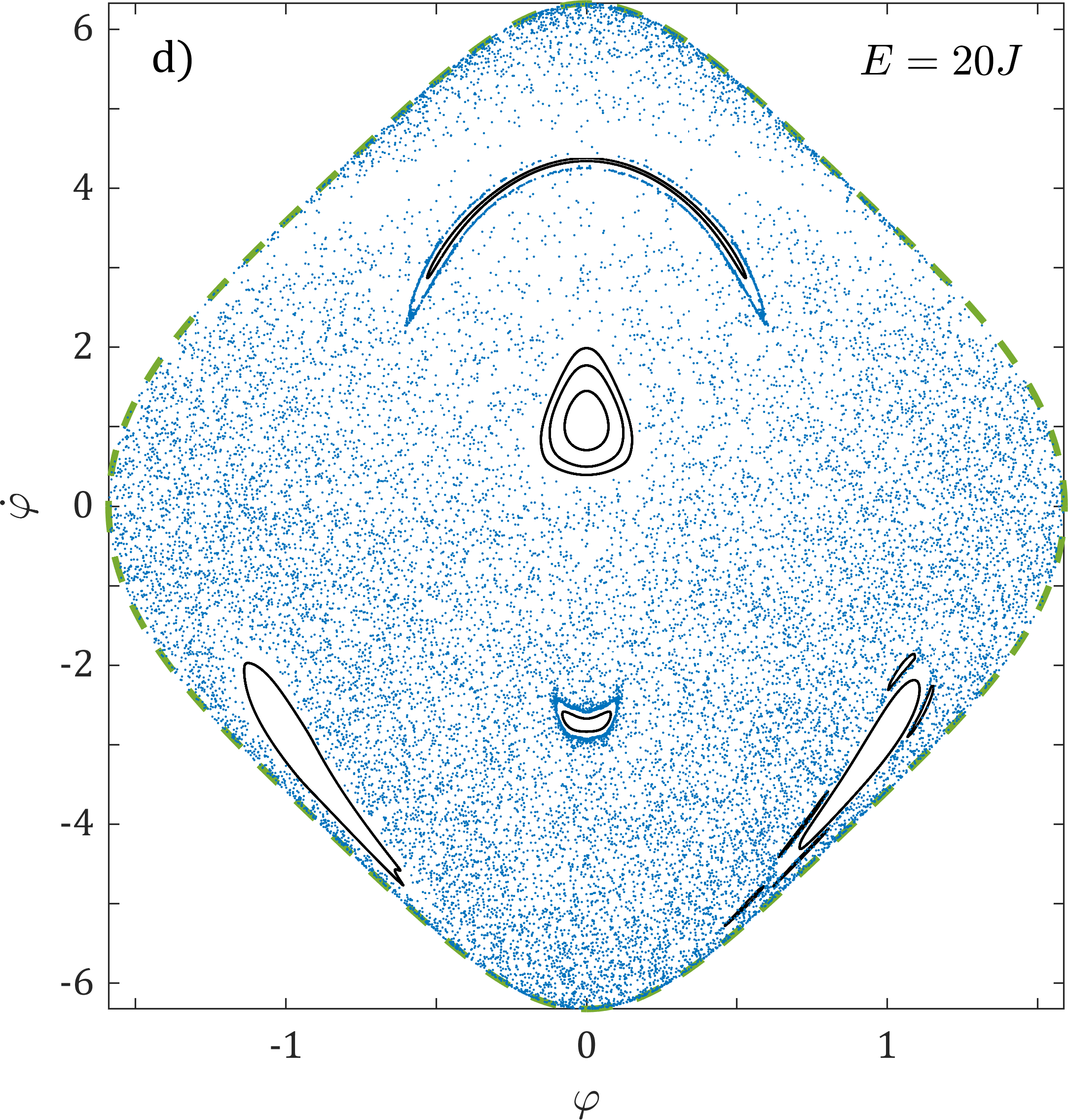}  
\includegraphics[width=0.45\textwidth]{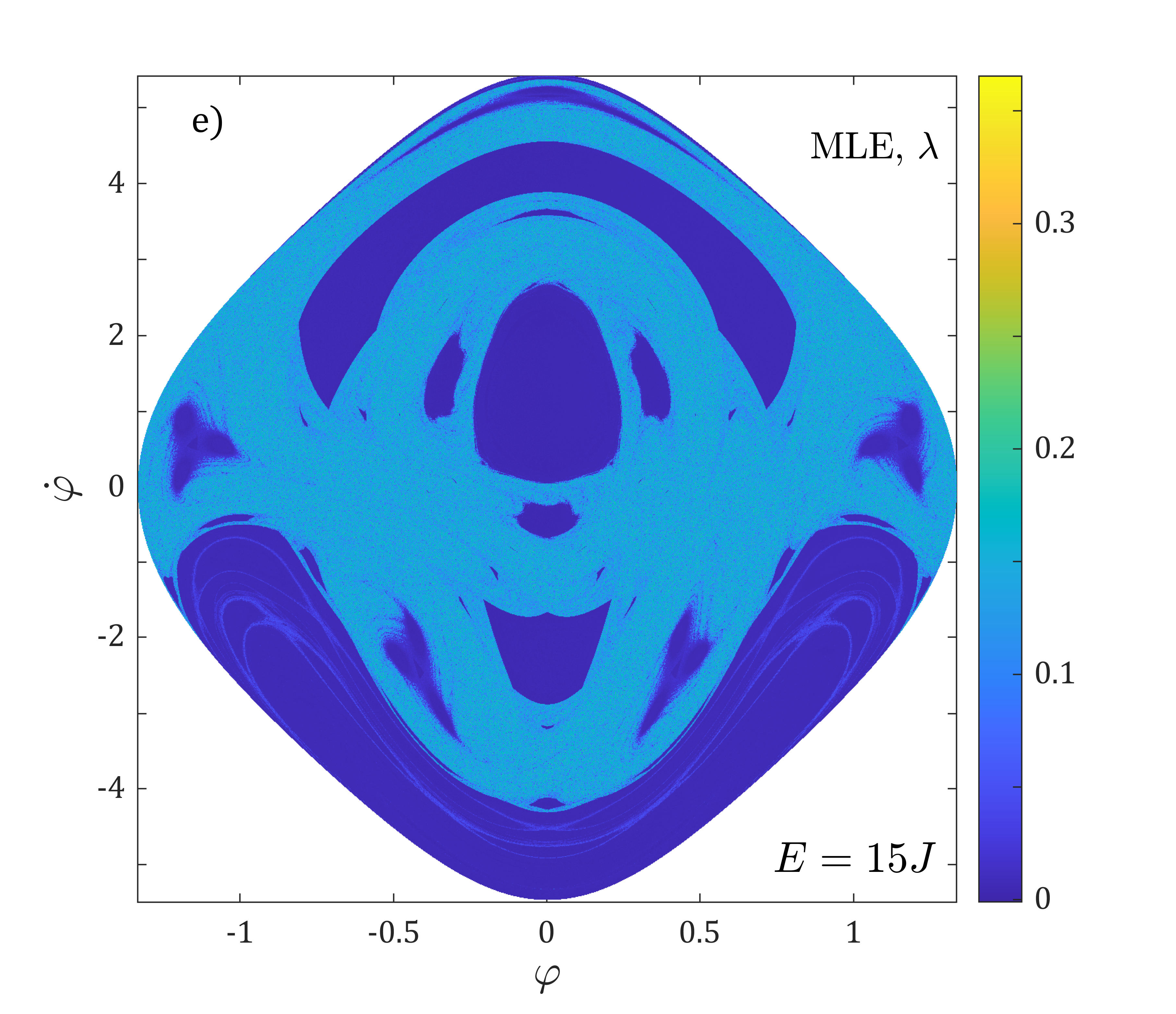}
\includegraphics[width=0.45\textwidth]{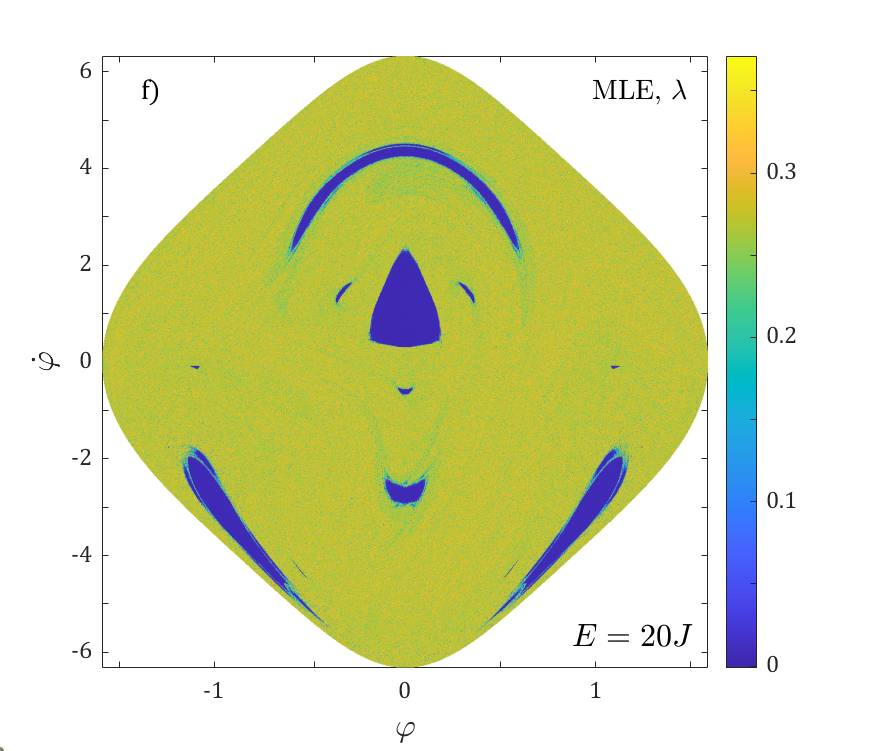}
\caption{Poincaré sections (a, b, c, and d) and Lyapunov exponent (e
  and f) for different energies indicated in each panel. We observe
  that as the energy increases, the regions covered by chaotic
  trajectories increase to the detriment of those corresponding to
  regular orbits. We note a clear correspondence between the values of
  positive (null) MLE and the chaotic (regular) regions. The MLE
  colormaps (e and f) comprise results calculated on a grid of $1200
  \times 1200$ initial conditions in the Poincaré section. }
\label{fig:variosT}
\end{figure*}

\subsection{Poincaré sections and MLE as a function of the energy}

To verify the performance of the method and to be able to analyze some
aspects of the motion, we obtained Poincaré sections for
representative values of the energy. To get these sections, we used
the numerical integrator we mentioned, \textit{ode45}, and the
intersection points were found with the integrator built-in function,
\textit{events}, which allows us to obtain specific topics of interest
of the trajectory. Precisely, we determine the points of the circuit
verifying conditions are given by
Eq.~\eqref{eq:definicionSecPoin}. The results are consistent with
those obtained in the literature
\cite{rafat2009dynamics,Chaso:3Rotors}.

Regular and chaotic trajectories are distinguishable in
Fig.~\ref{fig:variosT}: filling area points corresponding to chaotic
trajectories and filling curves points or isolated points
corresponding to quasiperiodic or periodic orbits.  In Fig.
\ref{fig:variosT}(a), for the given energy value the generic
trajectories are quasiperiodic (frequency ratio is irrational).  These
trajectories surround two regions where the normal oscillation modes
occur in the limit of small amplitudes.  In the subsequent panels, we
notice how these curves deform and decrease in size while chaotic
regions emerge as the energy increases. We also note that for an
energy of about $E=20J$, the chaotic region is already considerable,
and it can be seen that for an energy of about $E=30J$, regular
regions occur in a small region of the section.

Poincaré sections are compared with the MLE for energies of $E=15J$
and $E=20J$ to get a deeper insight. In Fig.~\ref{fig:variosT}(e) and
(f), we represent colormaps of the MLE. It is remarkable an excellent
agreement between panels (c) and (d), Poincaré sections, and (e) and
(f) MLE. Therefore, we conclude that both techniques provide coherent
results.

\subsection{Criterion for positive MLE}
\label{ssec:clasificacionLyapu}  

\begin{figure} [h]
\centering
\includegraphics[width=0.70\textwidth]{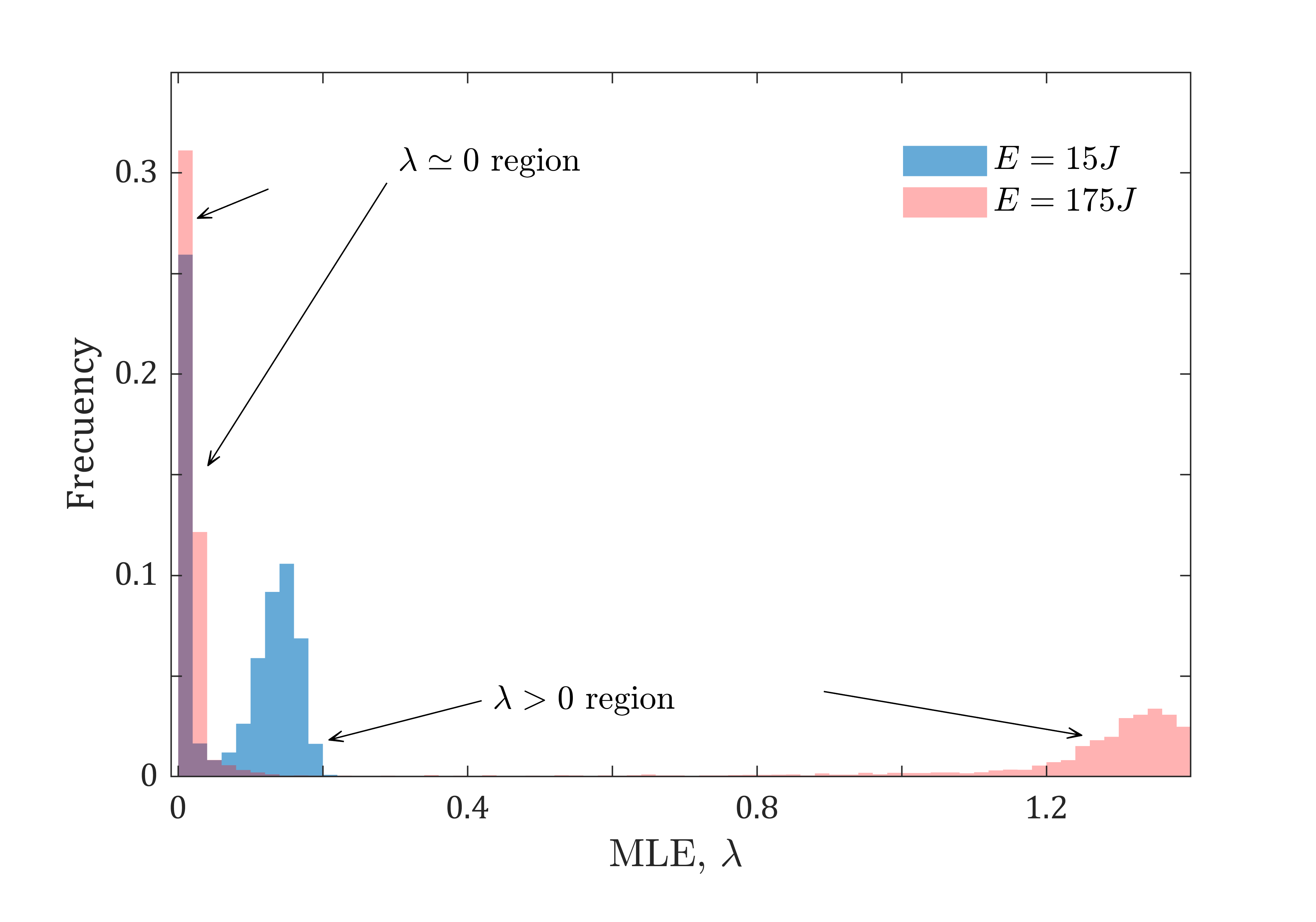}
\caption{MLE histogram for $E=15J$ and  $E=175J$ overlapped. Values are distributed in two regions:   the  $\lambda \sim 0$ region is associated with quasiperiodic orbits, while the $\lambda > 0$ region corresponds with chaotic orbits.}
\label{fig:ExpoLyapuHist} 
\end{figure}

One point that is worth exploring further is the criterion used to
classify the null or positive values of the MLE, $\lambda=0$ or
$\lambda> 0$, for each point in the last step of the algorithm
detailed in \ref{sec:alg}.  The chosen criterion is relevant because
it is based on estimating the limit of a series by a finite
sum. Hence, even if the limit is zero, it is not possible to obtain
precisely a null value in finite times.  To establish this criterion,
we studied the statistical distribution of MLE. In
Fig.~\ref{fig:ExpoLyapuHist} we can see a histogram of the
distribution of MLE varying the initial conditions but keeping the
energy fixed in two different values ($E=15$J or $E=175$J which
correspond to panels (c) and (d) of Fig.~\ref{fig:variosT}).  The
histograms reveal that although the MLE values do not take a single
value, we can clearly distinguish two clusters, one close to zero and
the other around a positive value.  Comparing to the Poincaré section,
we identify that the first cluster corresponds to $\lambda=0$ and the
second to $\lambda>0$.  This fact allows us to establish a threshold
to determine whether a calculated MLE value is zero or positive.

It is worth mentioning that the ranges of values corresponding to a
null MLE may vary. In general, we observe that by keeping the number
of iterations fixed and increasing the energy, the convergence is
slower, an event that leads to the MLE values of the stable regions
obtained being higher for higher energies.  We can appreciate this
phenomenon in Fig.~\ref{fig:ExpoLyapuHist}. There, we see how the
values corresponding to the $0$ region for an energy of $175J$ are
more significant than those of $15J$.  However, contrasting again with
the Poincaré sections, we can corroborate that the initial conditions
whose exponents fall within this first region of the histogram (for
energy $175J$) correspond to quasiperiodic trajectories, curves in the
Poincaré section, and the same with the points in the second region of
the histogram with the chaotic zones. Therefore, we adopt this
criterion to classify the MLE as null or positive.  We also note that
MLE values obtained are consistent with other studies on the subject
\cite{stachowiak2006numerical}, for example, values of the order of
$10^{-2}$ for low energies.

\subsection{Chaotic fraction in the Poincaré section}
\label{ssec:fraccionCaotica}

By obtaining and classifying the maximum Lyapunov exponents for the
grid points in the Poincaré sections, we can approximate the fraction
associated with chaotic or regular motions as
\begin{equation}
    \text{Chaotic fraction}=\frac{\text{Number of chaotic initial conditions}}{\text{Total number of initial conditions}}.
\end{equation}
the objective is the variation of this fraction as a function of the
system's mechanical energy. For this study, simulations were performed
on a grid of $100\times 100$ points of the Poincaré section, sweeping
energies between $1J$ and $2000J$. For each initial condition, a
reference orbit of $\SI{1000}{\second}$ was integrated, and $400$
initial conditions close to this were taken for the calculation of the
MLE, integrating each short period of $\SI{2.5}{\second}$ and then
resetting the close initial condition.

\begin{figure*}
\centering
\includegraphics[width=0.7\textwidth]{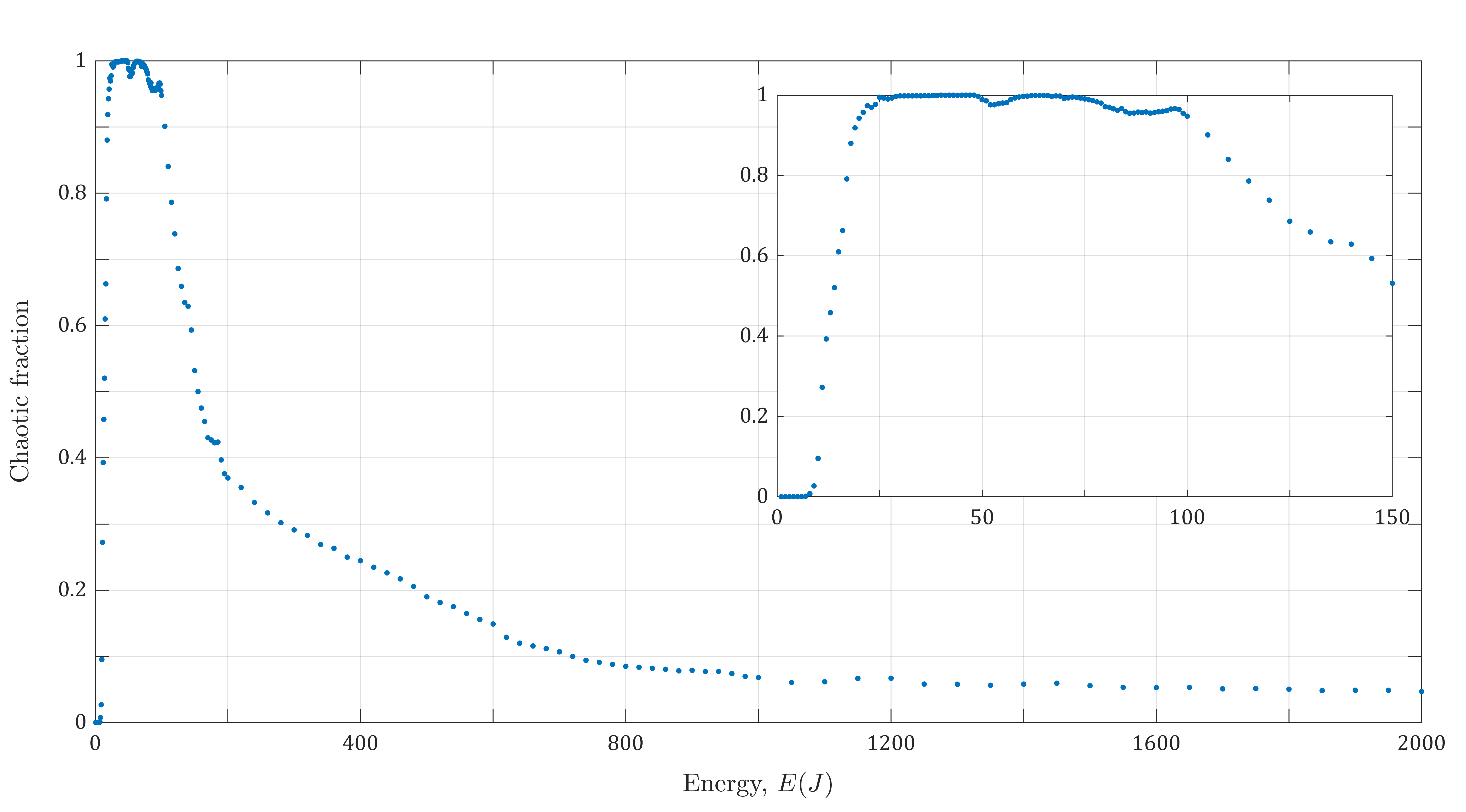}
\caption{Chaotic fraction of the system as a function of energy. On
  the upper right, we appreciate a detail of the results for low
  energies. We note how the initial value is zero, then grows with E
  until reaching the maximum value 1, and then decreases for high
  energies.}
\label{fig:espectroCompleto}
\end{figure*}

In Fig.~\ref{fig:espectroCompleto}, we can see the global behavior of
the chaotic fraction as a function of the energy. We note,
particularly in the inset corresponding low energies, that the chaotic
fraction is almost zero for energies going to zero. This limit
coincides with the region of validity of the small oscillations
approximation in which the general motion of the pendulum is
quasiperiodic and, therefore, non-chaotic. If we continue analyzing
the Fig.~ \ref{fig:espectroCompleto}, we see that as energy increases,
the chaotic fraction begins to increase in the same way, eventually
almost covering the entire Poincaré section for values greater than
$E=25J$. Then, as the energy continues to grow, we notice that, at
approximately $E=100J$, its value begins to decrease, and it does so
monotonically in the range of energies shown. Because of the
regularity of the system as the energy increases, as we approach the
integrable limit at high energies, we expect the monotonic decrease to
be robust. This behavior, a ``linearity $\rightarrow$ chaos
$\rightarrow$ linearity transition '' appears to be common to several
simple chaotic conservative systems, such as a set of rotors
\cite{Chaso:3Rotors} or a double pendulum formed by square plates
\cite{rafat2009dynamics}, and probably to many more, where in
particular we have integrable limits at the two extremes of the energy
range. This is one of the relevant conclusions of the work.

\begin{figure}
    \centering \includegraphics[scale=0.65]{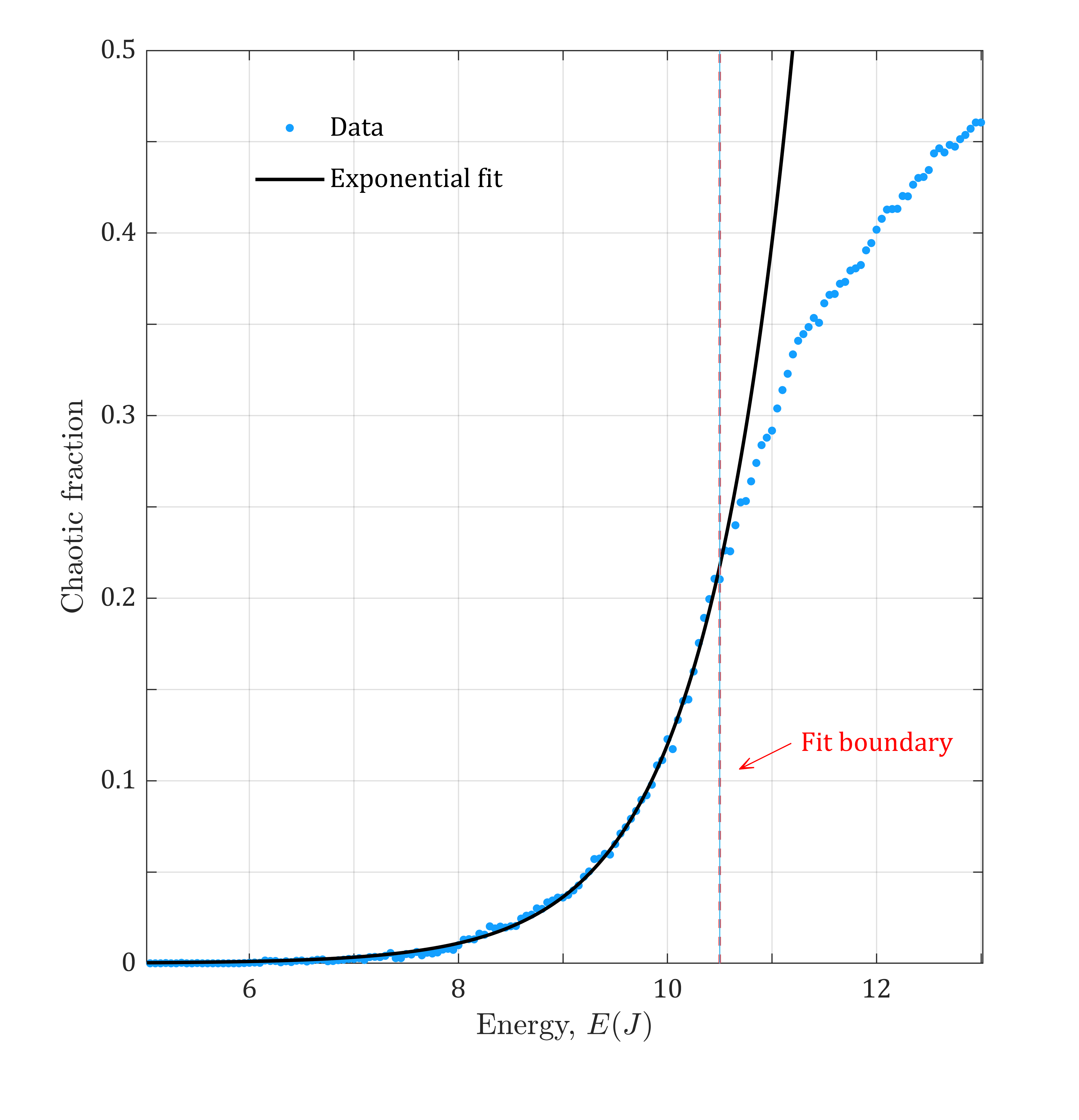}
    \caption{Chaotic fraction for low energies. On the numerically
      obtained data, an exponential fit is used. We can appreciate
      that the behavior fits very satisfactorily by the exponential,
      up to a value of $E=10.5J$ approximately.}
    \label{fig:espectroCorto}
\end{figure}

One issue that arises, mainly from analyzing the low energy region
depicted in Fig.~\ref{fig:espectroCompleto}, is the initial growth
rate of the chaotic fraction. The calculations were repeated in the
low energy regions, varying E from $0.05J$ to study this. The behavior
described by the chaotic fraction, for values of up to $E = 10.5$J, is
very satisfactorily fitted by an exponential curve. The coefficient of
determination for such a fit is $r^2=0.998$. Not surprisingly, the fit
is not appropriate for higher energies because the chaotic fraction,
bounded to its maximum value, 1, must decrease its slope. In
Fig.~\ref{fig:espectroCorto}, we can see the values obtained for the
chaotic fraction for low energies with the exponential fit. We can
appreciate an excellent agreement between the fit and the numerical
data for this energy range. It is interesting to ask how much we can
extrapolate this behavior to other systems. This raises a question for
future work on the subject. It would be exciting to analyze other
simple systems and study if they present this characteristic of
initial exponential growth of the chaotic fraction.

\section{Conclusion}
\label{sec:conclusiones}

In this paper, we studied the dynamics of a double-plane pendulum,
particularly the fraction of regular and chaotic trajectories as a
function of the system's mechanical energy. By analyzing Poincaré
sections, we could appreciate how, as the energy increases, the stable
regions of the phase space - those associated with quasiperiodic
trajectories - yield space to the chaotic ones - related to erratic
orbits, without any regularity. Eventually, the system's motion
reaches a state of global chaos - any initial condition given to it
leads to chaotic motion.

This transition was analyzed quantitatively using the numerical
calculation of the maximum Lyapunov exponent. Using a discretization
of the Poincaré section (through a grating), we could estimate the
chaotic fraction of the initial conditions in the phase space as the
quotient between those associated with chaotic orbits and the total
number for a given energy. In agreement with the theoretical
predictions we explained in \ref{ssec:general}, we could appreciate,
for example, in Fig.~\ref{fig:espectroCompleto}, how the system has a
high regularity for very low or very high energies. In both cases, the
system's motion is orderly; we may be in the Hamiltonian's integrable
limits that describe its evolution. On the other hand, we verify that
for an intermediate range of energies, essentially every initial
condition leads to a chaotic trajectory. Therefore, and as we expected
theoretically, the single, double pendulum undergoes an integrability
transition $\rightarrow$ chaos $\rightarrow$ integrability.

We focused more on the region of energies where the chaotic fraction
grows, starting from its null value in the limit of small
oscillations. We see the detail in Fig. \ref{fig:espectroCorto}. We
verified that in a range of energies up to about $10.5J$, the growth
of the chaotic fraction fits very well with an exponential curve.  For
future work, the most remarkable is this last mentioned result, the
exponential growth of the chaotic fraction for low energies.

\section*{Acknowledgments}

Most numerical experiments were performed using resources from
ClusterUY (https://cluster.uy), a national supercomputing center that
allows high-performance scientific computations. The simulations took
approximately 500 CPU hours with 20 Intel Xeon-Gold 6138
cores. E.D.L. acknowledges support from Brazilian agencies CNPq
(No. 301318/2019-0) and FAPESP (No. 2019/14038-6 and
No. 2021/09519-5).

\bibliographystyle{apsrev4-1}
\bibliography{BIBLIO}

\end{document}